\documentclass[referee]{aa} 
\usepackage{graphicx}
\usepackage{txfonts}
\usepackage{float}
\usepackage{latexsym}
\usepackage{color}
\usepackage{hyperref}

\def\k{km s$^{-1}$}
\def\jy{Jy beam$^{-1}$}
\def\ks{km s$^{-1}$}

\def\m{$^\prime$}
\def\s{$^{\prime\prime}$}

\def\cm3{cm$^{-3}$}
\def\cm2{cm$^{-2}$}

\def\2{$^{12}$CO}
\def\3{$^{13}$CO}
\def\msol{$M_\odot$}
\def\cchi{$\chi^{2}$}
\def\cchib{$\chi_{best}^{2}$}
\def\HII{H\,{\sc{ii}}}
\def\HI{H\,{\sc{i}}}

\begin{document}

\title{The infrared dust bubble N22: an expanding \HII\ region and the star formation around it}

\author {Wei-Guang Ji    \inst{1,2}
         \and Jian-Jun Zhou  \inst{1,3}
         \and Jarken Esimbek  \inst{1,3}
         \and Yue-Fang Wu  \inst{4}
         \and Gang Wu  \inst{1,3}
         \and Xin-Di Tang  \inst{1,2}
}
\institute{ Xinjiang Astronomical Observatory, Chinese Academy of Sciences, Urumqi 830011, PR China\\
             \email{jiweiguang@xao.ac.cn}
             \and Graduate University of the Chinese Academy of Sciences, Beijing 100080, PR China
             \and Key Laboratory of Radio Astronomy, Chinese Academy of Sciences, Urumqi 830011, PR China
             \and Department of Astronomy, Peking University, 100871, Beijing China
}


   \date{Accepted 30/05/2012}


  \abstract
    {} 
   { To increase the observational samples of
   star formation around expanding \HII\ regions,
   we analyzed the interstellar medium and star formation around N22. }
   { We used data extracted from the seven large-scale surveys from infrared to radio wavelengths.
   In addition we used the JCMT\thanks{James Clerk Maxwell Telescope~~~http://www.jach.hawaii.edu/JCMT/} observations of the $J=$ 3-2 line of \2 emission data
   released on CADC\thanks{The Canadian Astronomy Data Centre~~~http://www3.cadc-ccda.hia-iha.nrc-cnrc.gc.ca/} and the \2 $J=$ 2-1 and $J= $3-2 lines observed by the KOSMA\thanks{KOSMA~~~http://www.astro.uni-koeln.de/kosma/} 3 m telescope.
   We performed a multiwavelength study of bubble N22.}
   {A molecular shell composed of several clumps agrees very well with the border of N22,
   suggesting that its expansion is collecting the surrounding material.
   The high integrated \2 line intensity ratio $R_{I_{{\rm CO(3-2)}}/I_{{\rm CO(2-1)}}}$ (ranging from 0.7 to 1.14)
   implies that shocks have driven into the molecular clouds.
   We identify eleven possible O-type stars inside the \HII\ region,
   five of which are located in projection inside the cavity of the 20 cm radio continuum emission and
   are probably the exciting-star candidates of N22.
   Twenty-nine YSOs (young stellar objects) are distributed close to the dense cores of N22.
   We conclude that star formation is indeed active around N22;
   the formation of most of YSOs may have been triggered by the expanding of the \HII\ region.
   After comparing the dynamical age of N22 and the fragmentation time of the molecular shell,
   we suggest that radiation-driven compression of pre-existing dense clumps may be ongoing. }
   {}

   \maketitle
%

\section{Introduction}
\label{sec 1}

Using the Spitzer-GLIMPSE\footnote{Galactic Legacy Infrared Mid-Plane Survey Extraordinaire\\ http://irsa.ipac.caltech.edu/data/SPITZER/GLIMPSE/} survey of the Galactic plane (Benjamin et al.~\cite{ben03}),
Churchwell et al. (\cite{chu06, chu07}) cataloged almost 600 infrared (IR) dust bubbles.
The IR dust bubbles are bordered by bright 8 $\mu$m shells (the photo-dissociation region (PDR))
that encloses bright 24 $\mu$m interiors.
Of these IR dust bubbles, Deharveng et al. (\cite{deh10})
have studied a series of 102 ionized bubbles and concluded
that 86 \% of these bubbles enclose \HII\ regions.

Deharveng et al. (\cite{deh10}) also concluded that star formation triggered
by \HII\ regions may be an important process, especially for massive-star formation.
Elmegreen (\cite{elm98}) and Deharveng et al. (\cite{deh05})
proposed various physical mechanisms of triggered the formation of stars around the \HII\ regions.
Recently, one of the triggered processes called "collect and collapse",
proposed by Elmegreen \& Lada ~(\cite{elm77}), has been studied in more detail inside the \HII\ region borders,
and several observational studies supported that this mechanism is ongoing in several \HII\ regions
(see e.g. Zavagno et al.~\cite{zav07} \& \cite{zav10}; Pomar¨¨s et al.~\cite{pom09}; Petriella et al.~\cite{pet10}).

N22 is one of the northern IR dust bubbles cataloged by Churchwell et al. (\cite{chu06}).
By checking the infrared (8 and 24 $\mu$m) image and the JCMT \2 $J=$ 3-2 data of N22,
we found collected material, which indicates that triggered star formation may be taking place around N22.
In this work, we present a molecular and IR study of the environment surrounding the IR dust bubble N22
to explore its surrounding ISM and search for signatures of star formation.
We describe our data in Sect. 2,
present N22 in Sect. 3,
and the molecular analysis in Sect. 4 along with the exciting star(s) and star formation,
Sect. 5 details our analysis of the star formation mechanism,
and we summarize our findings in Sect.6.

\section{Data}
\label{sec 2}

Seven large-scale surveys were used in our study:
2MASS\footnote{Two Micron All Sky Survey~~~http://pegasus.phast.umass.edu/},
GLIMPSE, MIPSGAL\footnote{MIPSGAL~~~http://irsa.ipac.caltech.edu/data/SPITZER/MIPSGAL/},
NVSS\footnote{NRAO VLA Sky Survey~~~http://www.cv.nrao.edu/nvss/},
GRS\footnote{Galactic Ring Survey~~~http://www.bu.edu/galacticring/new\_index.htm},
AKARI-BSC\footnote{Far-Infrared Survey (FIS) for AKARI Bright Source Catalog\\ http://heasarc.gsfc.nasa.gov/W3Browse/all/akaribsc.html},
and the BGPS\footnote{The Bolocam Galactic Plane Survey~~~http://milkyway.colorado.edu/bgps/} catalog.

The GLIMPSE and MIPSGAL (Carey et al. ~\cite{car05}) surbeys were performed
using the {\it Spitzer} Space Telescope (Werner et al.~\cite{wer04}).
We used the mosaicked images from GLIMPSE and MIPSGAL acquired
by {\it Spitzer}-IRAC (3.6, 4.5, 5.8 and 8 $\mu$m) (Fazio et al.~\cite{faz04})
and the MIPS instrument (24 and 70 $\mu$m) (Rieke et al. ~\cite{rie04}), respectively.
IRAC has an angular resolution of between 1.\s5 and 1.\s9. The MIPSGAL resolution at 24 $\mu$m is 6\s.
The four far-infrared wavelengths of AKARI are centered at 65, 90, 140, and 160 $\mu$m,
and their angular resolutions are 26.\s8, 26.\s8, 44.\s2 and 44.\s2, respectively (Murakami et al. \cite{mur07}).
The BGPS is a 1.1mm continuum survey with an angular resolution of 33\s\ (Aguirre et al. \cite{agu11}).
We also used the GLIMPSE Point-Source Catalog (GPSC), AKARI-BSC (Yamamura et al. ~\cite{yam10}),
and the BGPS catalog (Rosolowsky et al. ~\cite{ros10}).

In addition, we used the 1.4 GHz radio continuum emission data extracted
from the NVSS with an angular resolution of about 45\s~(Condon et al.~\cite{con98}).
For the molecular analysis, we used the James Clerk Maxwell Telescope (JCMT)
observations of the $J=$ 3-2 line of \2 emission data released on CADC
and the \3 $J=$ 1-0 from GRS performed by the Boston University and the Five College Radio Astronomy Observatory (FCRAO).
The co-added spectral cubes of the JCMT \2 $J=$ 3-2 line were binned to 0.2 \k,
and the data were smoothed with a 6\s\ Gaussian kernel.
The final spatial resolution of each cube is 16\s\ (Smith et al.~\cite{smi08}),
while the angular and spectral resolution of the GRS \3 $J=$ 1-0 are 46\s\ and 0.2 \k\ (Jackson et al.~\cite{jac06}).

We observed the \2 $J=$ 2-1 and $J=$ 3-2 lines using the KOSMA 3 m telescope at Gornergrat, Switzerland in 2010 February.
The half-power beamwidths of the telescope at observing frequencies of 230.538 GHz and 345.789 GHz are 130\s\ and 80\s,
and their corresponding system temperatures are about 233.5 and 357.8 K.
The pointing and tracking accuracy is better than 10\s.

\section{Presentation of N22}
\label{sec 3}

\object{N22} is a complete IR dust bubble centered on $\alpha_{2000} =$ 276\fdg325, $\delta_{2000} = -$13\fdg176~
($l = 18$\fdg$254$, $b = -0$\fdg$305$) (Churchwell et al.~\cite{chu06})
with a radius of about 1.77~pc (Beaumont \& Williams ~\cite{bea10}).
N22 was also known as an \HII\ region (G$18.259-0.307$) (Kolpak et al. ~\cite{kol03}),
its hydrogen recombination line velocity $v_{\mathrm{LSR}}$ is $\sim$ 50.9 \ks.
Kolpack and collaborators derived the near distance by comparison with the \HI\ absorption data of G$18.259-0.307$,
and obtained kinematic distances of $\sim 4.1\pm0.3$ kpc.
We adopt their $v_{\mathrm{LSR}}$ and kinematic distance for N22.

\begin{figure*}[htb]
\tiny
\centering
\includegraphics[width=17cm]{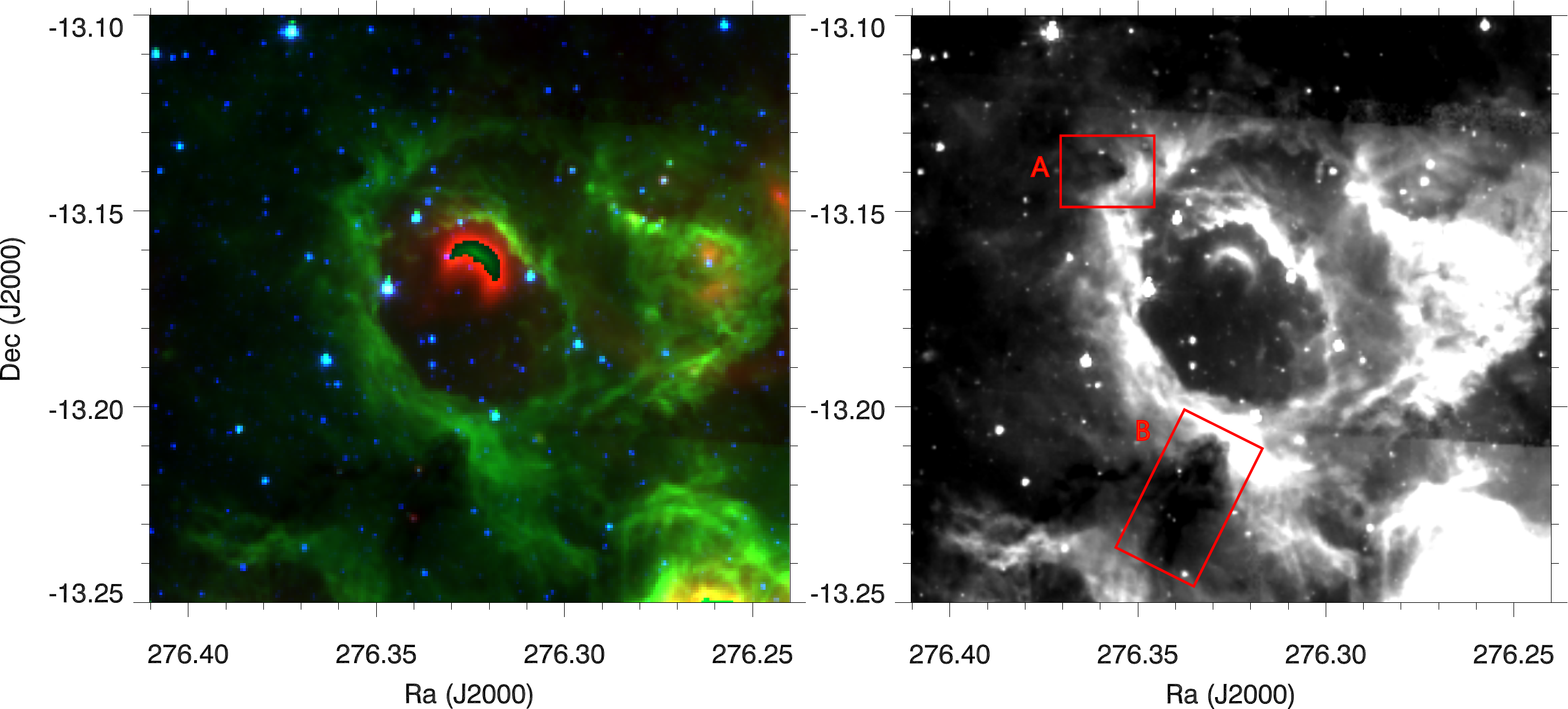}
\caption{Left: Mid-IR emission of the IR dust bubble N22: {\it Spitzer} three-color image (4.5 $\mu$m = blue, 8 $\mu$m = green, and 24 $\mu$m = red).
Right: The background shows the {\it Spitzer}-IRAC 8 $\mu$m emission. The two rectangles show the location of two IRDCs, which are named A and B.}
\label{1}
\end{figure*}

Fig. \ref{1} ({\it left}) shows a {\it Spitzer}-IRAC and {\it Spitzer}-MIPSGAL three-color image of N22
(4.5 $\mu$m in blue, 8 $\mu$m in green, and 24 $\mu$m in red).
The 24 $\mu$m emission corresponding to hot dust is distributed mainly toward the north of N22
(because of overexposure there are some bad data points in the peak of 24 $\mu$m ).
The PDR visible in the 8 $\mu$m emission is generally attributed to polycyclic aromatic hydrocarbon (PAH) molecules.
Infrared emission from PAHs is a good tracer of ionization fronts (IF)
because these large molecules are believed to be destroyed inside the ionized region,
but are excited in the PDR by the radiation leaking from the \HII\ region (Pomar¨¨s et al.~\cite{pom09}).
The absorption in 8 $\mu$m usually indicates infrared dark clouds (IRDCs).
Two absorptions appear in the gray map of 8 $\mu$m (right panel of Fig. \ref{1}), they were marked by two red rectangles.
Because these two absorptions are bordered by bright 8 $\mu$m and the ionization fronts are distorted,
we conclude that these two structures are indeed IRDCs (hereafter IRDC-A and IRDC-B).
They have not been cataloged in the IRDC catalogs of Simon et al. (\cite{sim06}) and Peretto \& Fuller (\cite{per09}).
However, IRDC-B has been identified by Watson et al. (\cite{wat08})
when they studied the nearby infrared dust bubble N21.

\begin{figure}[htb]
\tiny
\resizebox{\hsize}{!}{\includegraphics{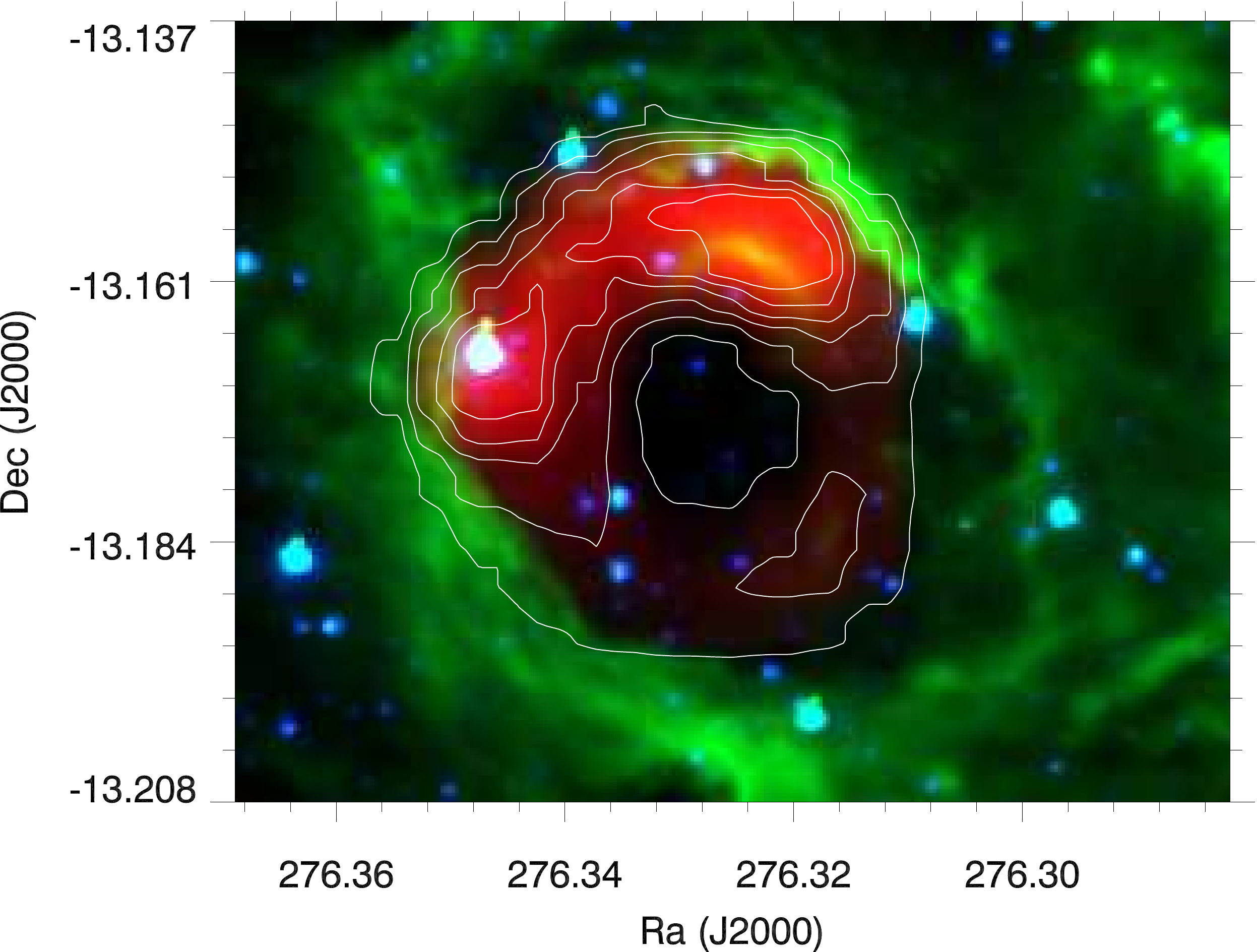}}
\caption{{\it Spitzer}-IRAC emission at 4.5 $\mu$m (in blue), 8 $\mu$m (in green), and the NVSS radio continuum emission at 20 cm (in red and emphasized with white contours). The levels range from 0.05 to 0.15 by 0.02 \jy.
The $\sigma_{\rm rms}$ of the NVSS data is 0.45 mJy beam$^{-1}$.}
\label{2}
\end{figure}

Fig. \ref{2} shows the {\it Spitzer}-IRAC emission at 4.5 $\mu$m (in blue), 8 $\mu$m (in green),
and the NVSS radio continuum emission at 20 cm (in red and emphasized with white contours).
The 20 cm emission, which is commonly attributed to the free-free emission from \HII\ regions,
is bounded by 8 $\mu$m emission.
The 20cm emission shows a shell morphology, and a cavity can be clearly seen at its center.
The strong 20 cm emission is distributed in the north of the shell, and shows an arc-like structure.
Two peaks appears in the arc, and the stronger one is coincident with a small arc traced by 8 $\mu$m emission.

\section{Molecular analysis and star formation}
\label{sec 4}

\subsection{Molecular analysis}
\label{sec 4.1}

\begin{figure}[htb]
\tiny
\resizebox{\hsize}{!}{\includegraphics{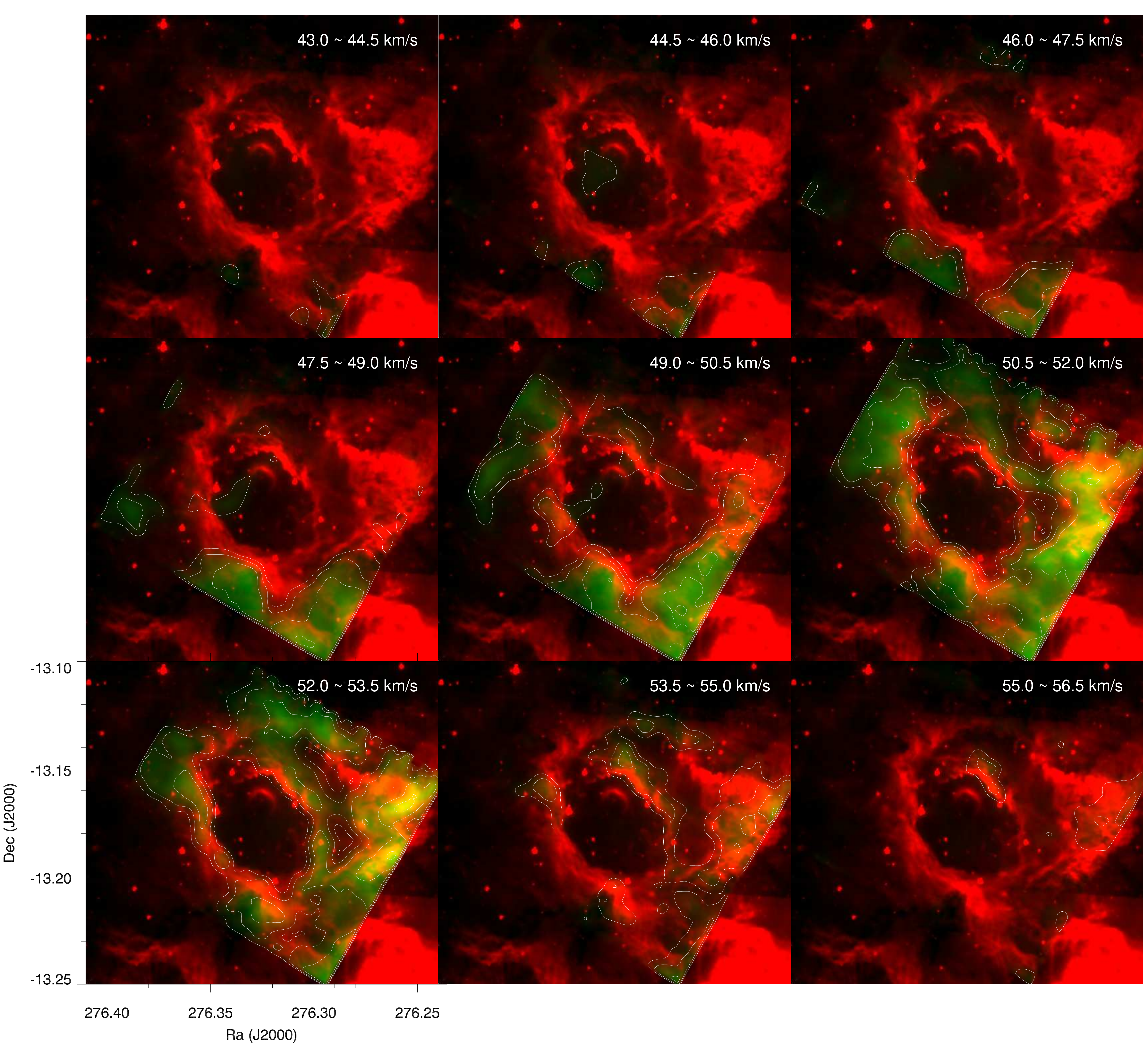}}
\caption{Integrated velocity channel maps of the JCMT \2 $J=$ 3-2 emission every 1.5 \ks\ from 43.0 to 56.5 \ks\ (in green) superimposed on the 8 $\mu$m emission (in red). The contour levels of the \2  $J =$ 3-2 emission are 5, 10, 20, 35, and 55 K \k. The angular resolution of the JCMT \2 $J=$ 3-2 emission is 16\s.}
\label{3}
\end{figure}

We used the \2 $J=$ 3-2 data cube from JCMT to analyze the molecular environment around N22. Fig. \ref{3} shows the integrated velocity channel maps of the \2 $J=$ 3-2 emission from 43.0 to 56.5 \k\ in steps of 1.5 \ks. Between 49.0 and 55.0 \k, several molecular clumps are distributed across the border of N22, which may indicate where the collect-and-collapse process could be taking place.

\begin{figure}[htb]
\tiny
\resizebox{\hsize}{!}{\includegraphics{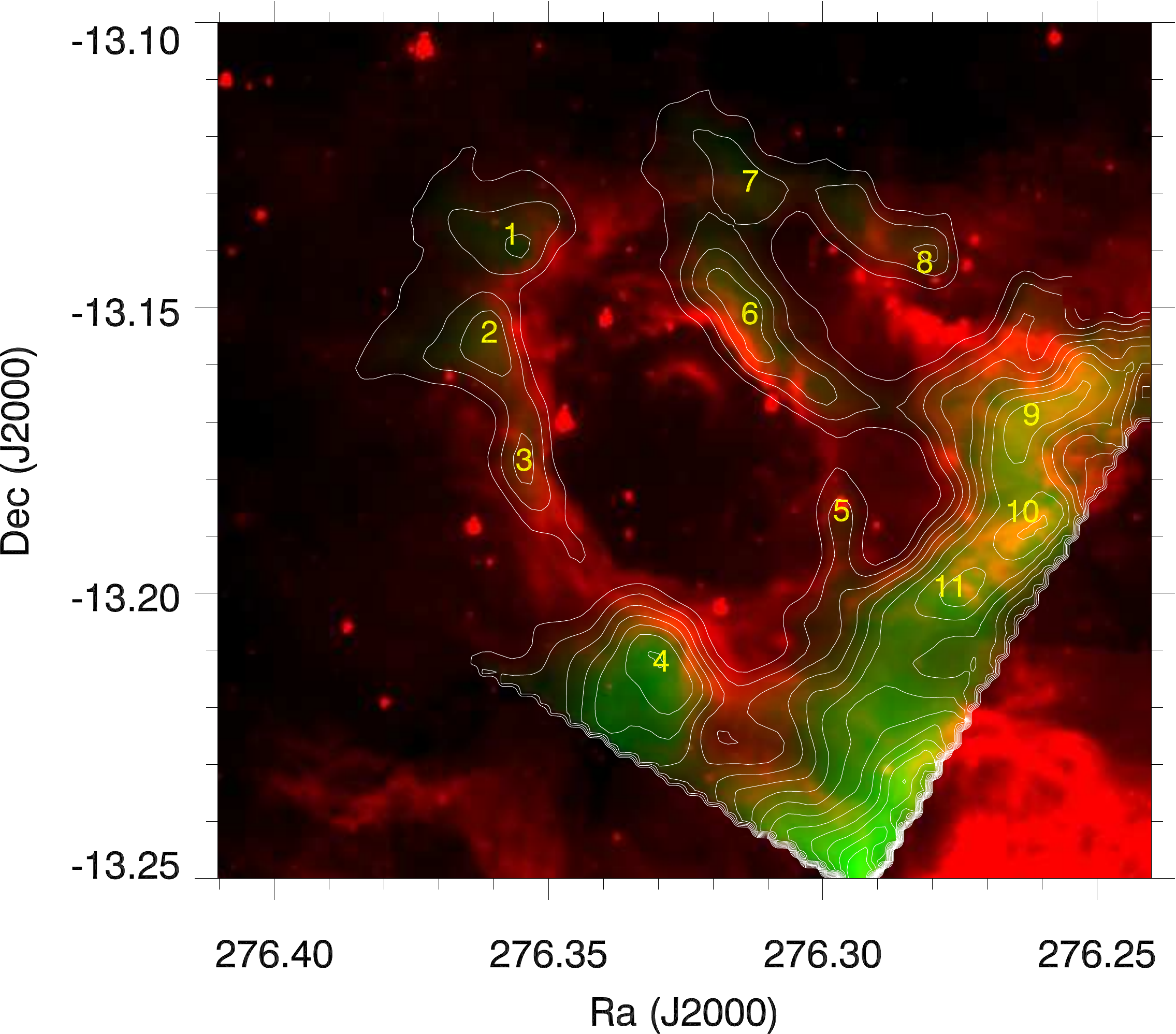}}
\caption{JCMT \2 $J=$ 3-2 emission (in green) integrated between 40 and 60 \ks. The contour levels of the \2 $J=$ 3-2 emission increase from 50.0 to 252.9 by 16.9 K \k. Red is the 8 $\mu$m emission. The yellow numbers are the ID of molecular cores.}
\label{4}
\end{figure}

We integrated the JCMT \2 $J=$ 3-2 emission between 40 and 60 \ks\ (in green) over the 8 $\mu$m emission (in red) (see Fig. \ref{4}). Molecular cores 1-11 are clearly identifiable and are indeed distributed across the borders of N22, which as pointed out by Deharveng et al. (\cite{deh05}), may be indicative of the collect-and-collapse mechanism.

From Fig.\ref{1} ({\it right}) and Fig.\ref{4}, we can see that the molecular cores 1 and 4 are associated with IRDC-A and IRDC-B.
These two IRDCs distort the IF excited at the 8 $\mu$m, which are presently compressed by the ionized gas. They may have preexisted there before the \HII\ region expanded to reach them. This is consistent with the model "radiation-driven compression" (Deharveng et al. \cite{deh10}). Based on the analysis above, we conclude that IRDC-B is physically associated with N22, and so is IRDC-A.

Assuming local thermodynamical equilibrium for the gas and an optically thick condition for the \2 $J=$ 3-2 line,
we estimated the excitation temperature of the \2 $J =$ 3-2 line following the equation $T_{{\rm ex}} = 16.6/\ln \left[1 + 1/\left(T_{{\rm mb}}/16.6 + 0.0024\right)\right]$ (Buckle et al.~\cite{buc10}), where $T_{{\rm mb}}$ is the main-beam temperature.
Because it is difficult to obtain the optical depth $\tau$ from one transition of CO, we used the $X$ factor between ${\rm H_{2}}$ and \2 to estimate the column density of the 11 molecular cores.
According to Shetty et al. (\cite{she11}), the $X$ factor is $\sim 1-6 \times 10^{20}$ \cm2 K$^{-1}$ km$^{-1}$ s.
Here we assume $X = 6 \times 10^{20}$ \cm2 K$^{-1}$ km$^{-1}$ s and estimated the column density using the formula
$$N_{\rm H_{2}} = 6 \times 10^{20} W_{\rm CO}~[{\rm cm}^{-2}], $$
where $W_{\rm CO}$ is the observed \2 intensity, estimated following the equation $W_{\rm CO} = \int T_{{\rm mb}}\,d\nu~{\rm cm}^{-2}$ K \k.
If the cloud cores are approximately spherical in shape, the mean ${\rm H_{2}}$ number density is $n_{{\rm H_{2}}} = 1.62 \times 10^{-19} N_{H_{2}}/L$, where $L$ is the cloud core diameter in parsecs (pc).
The mass is given by $M_{{\rm H_{2}}} = 1.13 \times 10^{-4} \mu_{{\rm g}} m({\rm H_{2}}) D^2 S N_{{\rm H_{2}}}$ [\msol] (Buckle et al.~\cite{buc10}),
where $D$ is the distance in pc, $S$ is the pixel area in $arcsec^{2}$, $\mu_{{\rm g}} =$ 1.36 is the mean atomic weight of the gas, and $m({\rm H_{2}})$ is the mass of a hydrogen molecule.
Finally, we list $L$, $T_{{\rm ex}}$, $N_{{\rm H_{2}}}$, $M_{{\rm H_{2}}}$, and $n_{{\rm H_{2}}}$ of molecule cores 1 $\sim$ 11 in Cols. 5-9 of Table \ref{t1}, respectively. We used Gaussian fittings to obtain the peak temperature, velocity width, and central velocity.
They are listed in Cols. 2-4 of Table \ref{t1}.

\begin{table*}
\tiny
\centering
\caption{The parameters of the molecular cores 1-11}
\label{t1}
\begin{tabular}{ccccccccc}
\hline\hline
 core  &   peak temperature &  central velocity  &  velocity width   &    $ L $   &   $T_{{\rm ex}}$    &  $N_{{\rm H_{2}}}$  & $M_{{\rm H_{2}}}$  & $n_{{\rm H_{2}}}$  \\
       &         (K)        &      (\ks)         &      (\ks)        &    (pc)    &       (K)           &  (10$^{22}$ \cm2)   &    (10$^3$ \msol)  & (10$^3$ cm$^{-3}$) \\
\hline
  1    &        20.97       &       51.28        &      2.32         &    1.78    &      28.50          &         3.11        &        1.70        &       2.83         \\
  2    &        20.15       &       51.43        &      2.34         &    1.49    &      27.66          &         3.01        &        1.15        &       3.28         \\
  3    &        15.79       &       50.49        &      1.48         &    1.18    &      23.14          &         1.49        &        0.36        &       2.04         \\
  4    &        20.69       &       48.65        &      3.03         &    1.86    &      28.21          &         4.01        &        2.38        &       3.49         \\
  5    &        11.27       &       52.73        &      1.86         &    0.88    &      18.37          &         1.34        &        0.18        &       2.47         \\
  6    &        10.54       &       53.83        &      4.24         &    1.47    &      17.59          &         2.86        &        1.06        &       3.14         \\
  7    &        19.37       &       52.56        &      1.96         &    1.73    &      26.86          &         2.42        &        1.25        &       2.27         \\
  8    &        23.17       &       51.75        &      1.96         &    1.29    &      30.76          &         2.90        &        0.82        &       3.65         \\
  9    &        26.32       &       51.76        &      3.00         &    1.73    &      33.99          &         5.05        &        2.59        &       4.73         \\
 10    &        39.77       &       51.81        &      3.40         &    1.18    &      47.63          &         8.64        &        2.08        &      11.85         \\
 11    &        30.45       &       51.74        &      3.41         &    1.46    &      38.19          &         6.63        &        2.44        &       7.35         \\
\hline
\end{tabular}
\tablefoot{Comparing the spectra between \2 $J=$ 2-1 and \3 $J=$ 1-0 and channel maps of \2 $J=$ 2-1, we suggest that the dips in the 12CO 3-2 profiles of molecular cores \#2, \#4, \#6, and \#11 do not indicate self-absorption. Here we just consider the spectra main component associated to each clump of N22.}
\end{table*}

\begin{figure*}[htb]
\tiny
\centering
\includegraphics[width=17cm]{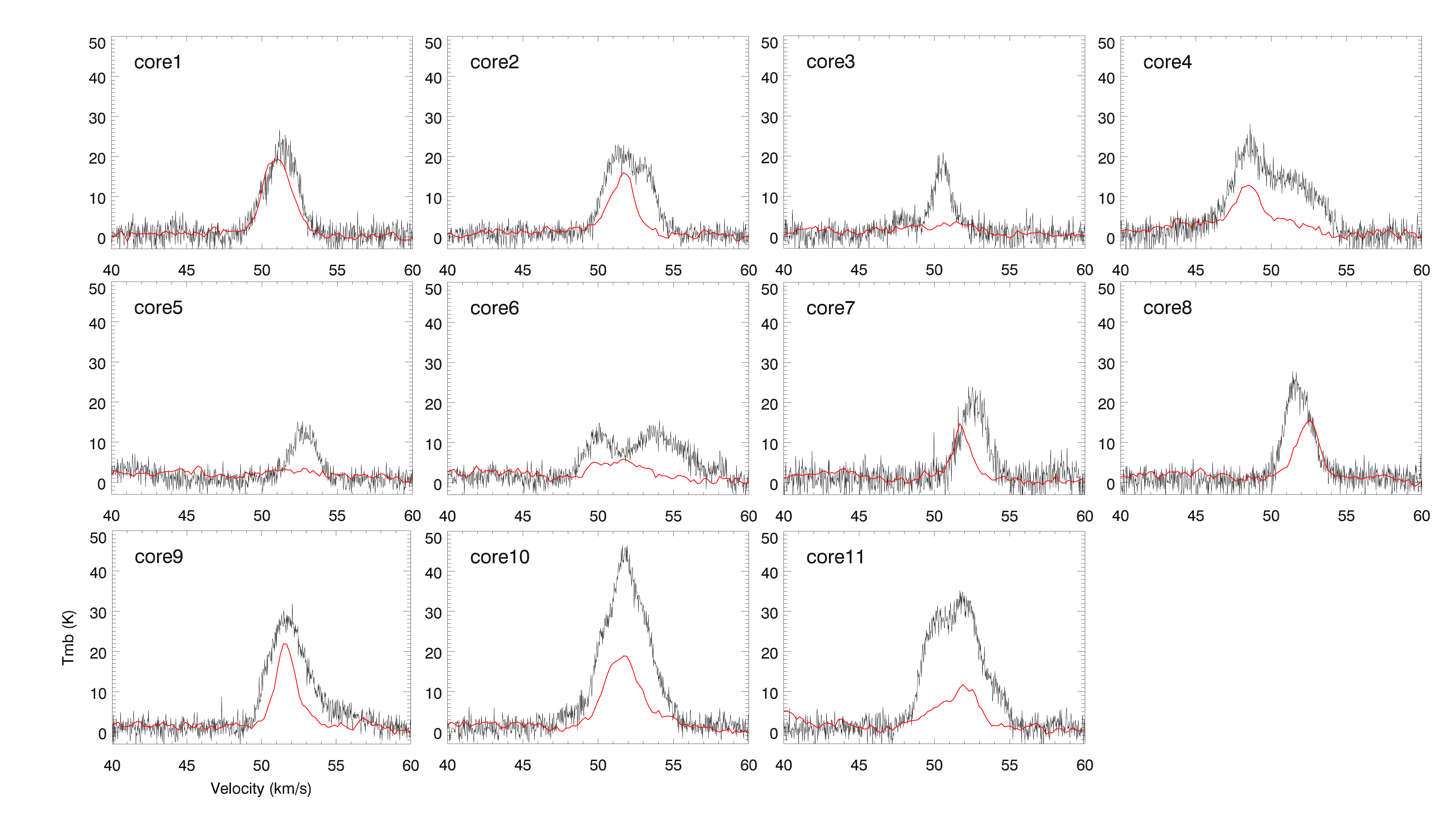}
\caption{JCMT \2 $J=$ 3-2 spectra at the peak of cores 1-11 are given in black, and GRS \3 $J=$ 1-0 spectra enlarged three times at the same region are delineated in red.}
\label{5}
\end{figure*}

Fig. \ref{5} displays spectra at the peak of the molecule cores 1-11,
the JCMT \2 $J=$ 3-2 spectra and GRS \3 $J=$ 1-0 spectra enlarged
three times are depicted in black and red. The \2 $J=$ 3-2
spectra of \#2, \#4, \#6, and \#11  seem to show self-absorption,
but corresponding \3 $J=$ 1-0 show no evidence of
self-absorption. Taking into consider the integrated velocity
channel maps (see Fig. \ref{3}), we suggest that \2 $J=$ 3-2 spectra
of \#2, \#4, \#6, and \#11 include some different velocity components.
Here we just consider the spectra main component associated to each clump of N22,
which is centered at the velocity quoted in Table \ref{t1}.

\subsection{The expanding \HII\ region interacting with its surrounding material}
\label{sec 4.2}

\begin{figure}[htb]
\tiny
\resizebox{\hsize}{!}{\includegraphics{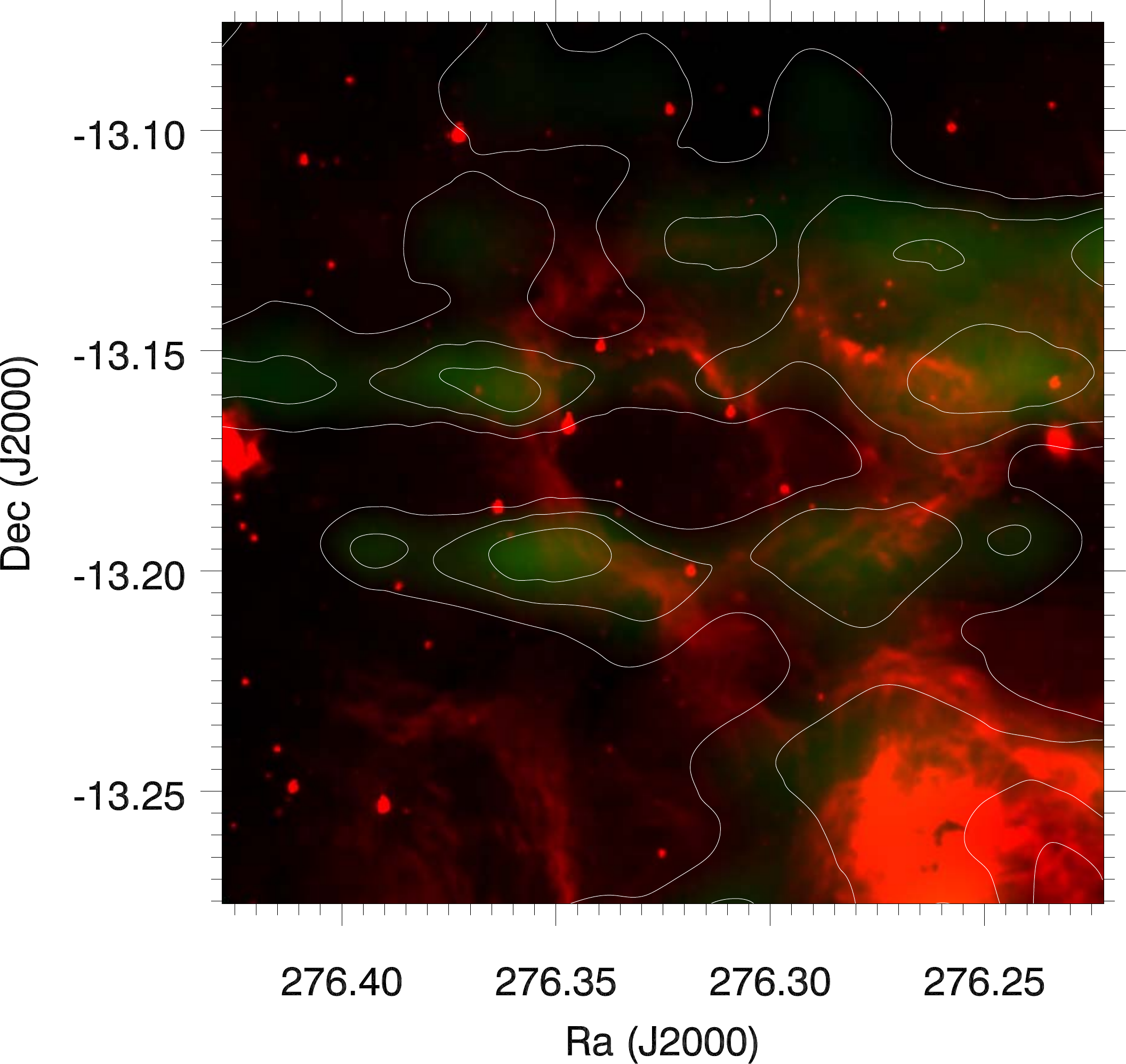}}
\caption{Line intensity ratios $R_{I_{{\rm CO(3-2)}}/I_{{\rm CO(2-1)}}}$ from KOSMA \2 $J=$ 3-2 and \2 $J=$ 2-1 is given in green, which is superimposed on the 8 $\mu$m emission in red. The contours range from 0.72 to 1.14 by 0.14.}
\label{6}
\end{figure}

As mentioned above, the morphology of N22 strongly suggests that the observed molecular shell has been swept and shaped by the expansion of N22 (see Fig. \ref{4}).
The different transitions of \2 can trace different molecular environments. To obtain the integrated intensity ratio of \2 $J=$ 3-2 to \2 $J=$ 2-1 ($R_{I_{{\rm CO(3-2)}}/I_{{\rm CO(2-1)}}}$), we convolved the \2 $J=$ 3-2 data to the same angular resolution as for \2 $J=$ 2-1.
We performed an integrated \2 line intensity ratio $R_{I_{{\rm CO(3-2)}}/I_{{\rm CO(2-1)}}}$ for the whole bubble using the KOSMA data, which is between 0.7 and 1.14 (see Fig. \ref{6}). The ratios around N22 are higher than 0.8, which is much higher than previous measurements of individual Galactic Molecular Clouds (MCs) (0.55, Sanders et al.~\cite{san93}) and even higher than the value (0.8) in the starburst galaxies M82 (Guesten et al.~\cite{gue93}). The higher line ratios imply that shocks have driven into the MCs (Xu et al.~\cite{xuj11}). Hereafter, the regions with higher line ratios ($> 0.8$) are called "the active regions". Our result indicates that the expanding \HII\ region is interacting with its surrounding material.

\subsection{Exciting stars}
\label{sec 4.3}

No exciting stars of N22 have been described in the literature. However,
according to the 20 cm flux and the estimated ionizing photon rate
of N22, Beaumont \& Williams (\cite{bea10}) stated that at least
ten O9.5 stars are needed to produce the observed radio flux. To
search for the exciting stars, we performed a photometric study of the
infrared point sources inside the \HII\ region based on the GLIMPSE
I Spring\m07 and the 2MASS All-Sky Point Source Catalogs.

\begin{table*}
\tiny
\caption{Exciting star candidates inside the \HII\ region}
\label{t2}
\centering
\begin{tabular}{cccccccccc}
\hline\hline
\# & GLIMPSE Desig.  & $J$  & $H$  & $Ks$  & $Av$  & $M_J$  & $M_H$& $M_{Ks}$  & O-type star  \\
\hline
 1 & G018.2516-00.3068 &  12.20  &  11.04 & 10.56  &  10.50  &  -3.83   &  -3.86   &  -3.68   &   O8V-O7V     \\
 2 & G018.2714-00.3026 &  12.18  &   9.50 &  8.18  &  24.29  &  -7.73   &  -7.81   &  -7.60   &   --          \\
 3 & G018.2443-00.3096 &  11.14  &   9.62 &  8.94  &  13.85  &  -5.83   &  -5.87   &  -5.68   &   --          \\
 4 & G018.2518-00.3097 &  14.49  &  12.05 & 10.97  &  21.30  &  -4.58   &  -4.74   &  -4.48   &   O5.5V-O4V   \\
 5 & G018.2668-00.2907 &  13.01  &  11.42 & 10.68  &  14.61  &  -4.17   &  -4.21   &  -4.02   &   O7V-O6V     \\
 6 & G018.2496-00.2955 &  13.60  &  12.25 & 11.65  &  12.40  &  -2.96   &  -2.99   &  -2.81   &   --          \\
 7 & G018.2559-00.3008 &  16.16  &  13.27 & 12.00  &  24.91  &  -3.93   &  -4.15   &  -3.86   &   O7.5V-O6V   \\
 8 & G018.2618-00.3044 &  13.93  &  11.10 &  9.81  &  24.75  &  -6.12   &  -6.30   &  -6.03   &   --          \\
 9 & G018.2627-00.3137 &  16.04  &  13.21 & 11.94  &  24.64  &  -3.97   &  -4.16   &  -3.89   &   O7.5V-O6V   \\
10 & G018.2532-00.3059 &  14.27  &  11.82 & 10.72  &  21.46  &  -4.85   &  -5.00   &  -4.75   &   O5V-O3V     \\
11 & G018.2660-00.2987 &  10.30  &   9.77 &  9.47  &   6.12  &  -4.50   &  -4.36   &  -4.28   &   O6V-O5V     \\
12 & G018.2680-00.3114 &  14.75  &  12.96 & 12.12  &  16.38  &  -2.93   &  -2.97   &  -2.78   &   --          \\
13 & G018.2521-00.3020 &  12.07  &  10.95 & 10.40  &  10.87  &  -4.06   &  -4.02   &  -3.88   &   O7V-O6.5V   \\
14 & G018.2441-00.2966 &  16.06  &  12.26 & 10.47  &  33.30  &  -6.40   &  -6.63   &  -6.33   &   --          \\
15 & G018.2531-00.2907 &  12.69  &  11.50 & 11.06  &  10.32  &  -3.29   &  -3.37   &  -3.16   &   --          \\
16 & G018.2380-00.3001 &  13.53  &  11.13 &  9.92  &  22.07  &  -5.77   &  -5.80   &  -5.62   &   --          \\
17 & G018.2431-00.3158 &  13.87  &  12.37 & 11.57  &  14.63  &  -3.32   &  -3.26   &  -3.13   &   --          \\
18 & G018.2456-00.3205 &  12.38  &  10.98 & 10.35  &  12.82  &  -4.30   &  -4.33   &  -4.15   &   O6.5V-O5.5V \\
19 & G018.2365-00.2990 &  16.04  &  12.07 &  9.99  &  36.36  &  -7.28   &  -7.35   &  -7.14   &   --          \\
20 & G018.2366-00.3083 &  14.67  &  12.06 & 10.88  &  22.80  &  -4.83   &  -5.00   &  -4.73   &   O5V-O3V     \\
21 & G018.2412-00.3181 &  10.19  &   9.58 &  9.39  &   5.68  &  -4.47   &  -4.48   &  -4.31   &   O6V-O5V     \\
22 & G018.2455-00.3170 &  14.37  &  11.78 & 10.52  &  23.45  &  -5.31   &  -5.39   &  -5.17   &   --          \\
23 & G018.2520-00.3115 &  13.93  &  12.49 & 11.86  &  13.01  &  -2.80   &  -2.85   &  -2.66   &   --          \\
\hline
\end{tabular}
\tablefoot{The numbers of the sources correspond to the numeration in Fig. \ref{7}.}
\end{table*}

\begin{figure}[htb]
\tiny
\resizebox{\hsize}{!}{\includegraphics{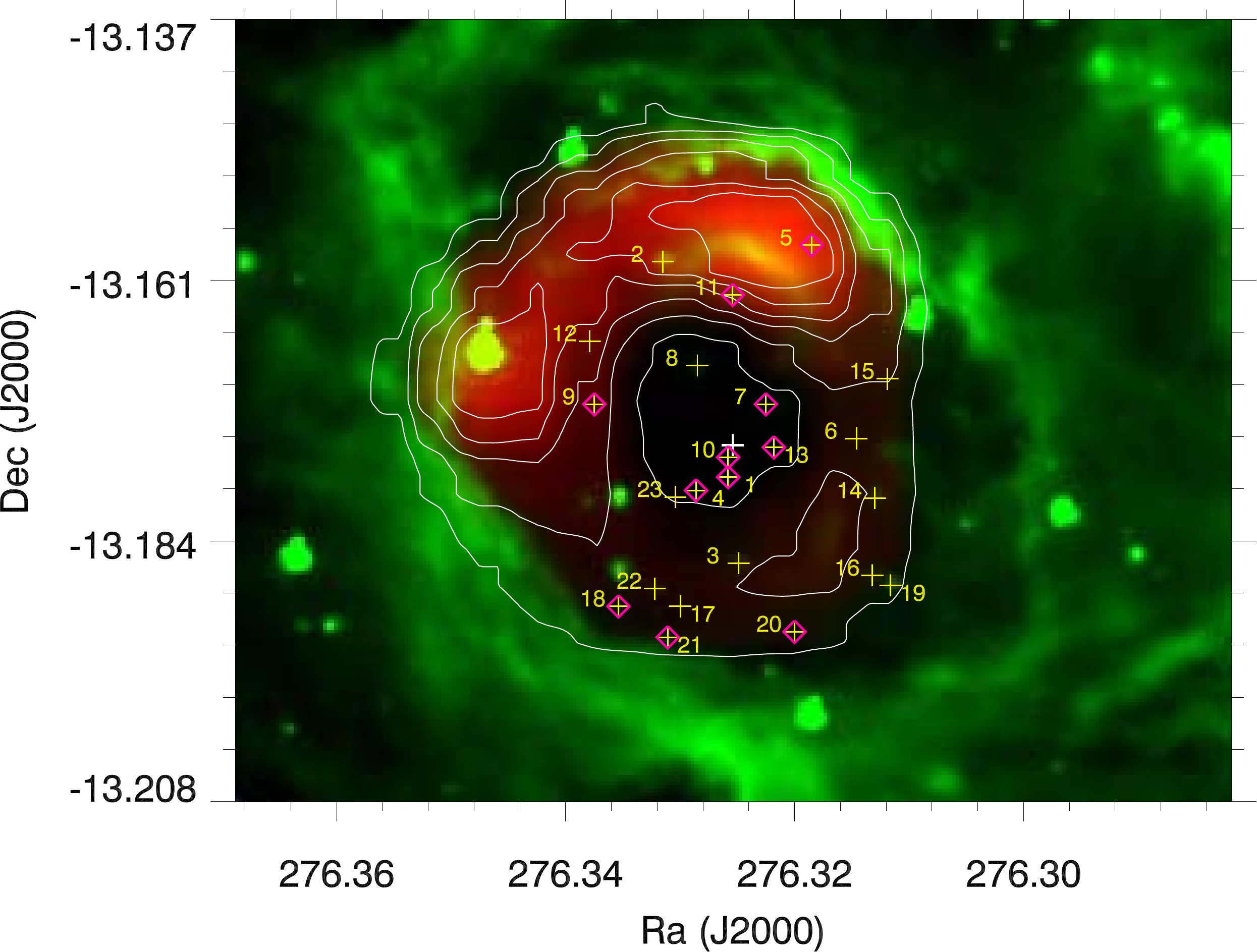}}
\caption{Two color image, 8 $\mu$m in red and 20 cm in green with contours,which are similar to Fig. \ref{2}.
The yellow plus show the location of the main sequence stars inside the \HII\ region.
The red diamonds are the O-type stars.
The white plus is the center of N22.}
\label{7}
\end{figure}

Taking into account that the exciting stars are expected to be in a
PAHs hole, we just considered the sources inside the \HII\ region.
They should be detected in the four {\it Spitzer}-IRAC bands and
three 2MASS bands. Finally, we found 24 sources. Following the color
criteria of Allen et al. (\cite{all04}), we found 23 main-sequence
(Class III) stars (see Fig. \ref{7}). Most of them are located inside the \HII\
region where the 20 cm radio continuum emission is weak.

To search for O-type stars, we used the $J$, $H$, and $K$ apparent
magnitudes obtained from the 2MASS Point Source Catalog to derive
the absolute {\it JHK} magnitudes. We assumed a distance of 4.1 kpc
and obtained the extinction for each source from the {\it (J-H)} vs.
{\it (H-K)} color-color diagram.
We also assumed the interstellar reddening law of
Rieke \& Lebofsky (\cite{rie85}) ($A_J/A_V$=0.282; $A_H/A_V$=0.175
and $A_K/A_V$=0.112) and the intrinsic colors $(J-H)_0$ and
$(H-K)_0$ obtained from Martins \& Plez (\cite{mar06}). By comparing
the derived absolute {\it JHK} magnitudes with those tabulated by
Martins \& Plez (\cite{mar06}), we found 11 O-type stars inside the
\HII\ region, these are \#1, \#4, \#5, \#7, \#9, \#10, \#11, \#13,
\#19, \#21, and \#22. Table \ref{t2} presents these sources with
their 2MASS designation (Col. 2), apparent {\it JHK} magnitudes
(Cols. 3-5), estimated extinctions (Col. 6), calculated absolute
{\it JHK} magnitudes (Cols. 7-9), and indicates whether their derived spectral
type coincides with an O-type in Col. 10.

\begin{table*}
\tiny
\caption{Near- and mid-IR fluxes of the 32 sources satisfying the photometric criteria of Allen et al. (2004) and the color criterion of $m_{4.5}-m_{8.0}\geq1$ around N22}
\label{t3}
\centering
\begin{tabular}{cccccccccc}
\hline\hline
Source & GLIMPSE Desig. &  $J$  &  $H$  & $K_{S}$ &  3.6 $\mu$m & 4.5 $\mu$m  &  5.8 $\mu$m &  8.0 $\mu$m & 24 $\mu$m \\
       &                & (mag) & (mag) &  (mag)  &    (mag)    &    (mag)    &     (mag)   &     (mag)   &   (mag) \\
\hline
YSO-1     &      G018.3251-00.3133   &     15.206   &   14.274   &   13.681   &    12.331  &    11.716  &    11.169   &   10.524 &   5.19  \\
YSO-2     &      G018.3145-00.3141   &              &            &   13.108   &    11.341  &    10.993  &    10.459   &    9.993 &         \\
YSO-3     &      G018.2351-00.3532   &     13.427   &   13.002   &   12.572   &    11.090  &     9.835  &     8.988   &    8.330 &   3.88  \\
YSO-4     &      G018.1992-00.3520   &              &            &   13.624   &    11.033  &    10.017  &     9.224   &    8.743 &   5.15  \\
YSO-5     &      G018.1997-00.3381   &              &   14.356   &   12.547   &    11.369  &    11.112  &    10.724   &   10.064 &   3.00  \\
YSO-6     &      G018.2262-00.3348   &              &            &            &    12.367  &    10.171  &     8.720   &    7.806 &   2.71  \\
YSO-7     &      G018.2171-00.3426   &              &            &            &    13.550  &    11.061  &     9.845   &    8.918 &         \\
YSO-8     &      G018.2249-00.3352   &              &            &   12.869   &    11.573  &    10.592  &     9.925   &    9.405 &         \\
YSO-9     &      G018.2157-00.3419   &              &            &            &    12.930  &    10.305  &     9.258   &    8.445 &         \\
YSO-10    &      G018.2277-00.3303   &              &            &            &    12.693  &    11.712  &    10.928   &   10.356 &         \\
YSO-11    &      G018.2149-00.2872   &              &            &            &    12.555  &    12.136  &    10.962   &   10.172 &   3.67  \\
YSO-12    &      G018.1976-00.2720   &     14.253   &   13.243   &   12.064   &    10.728  &    10.584  &    10.037   &    9.420 &   3.65  \\
YSO-13    &      G018.2031-00.2641   &              &   14.616   &   13.578   &    11.920  &    11.374  &    10.729   &   10.304 &         \\
YSO-14    &      G018.2307-00.2427   &              &            &            &    10.470  &     9.884  &     9.038   &    8.612 &   0.92  \\
YSO-15    &      G018.2318-00.2474   &              &            &   13.321   &    11.331  &    10.667  &    10.049   &    9.304 &   0.93  \\
YSO-16    &      G018.2336-00.2373   &              &            &            &    10.655  &     9.827  &     8.909   &    8.429 &         \\
YSO-17    &      G018.3058-00.2646   &              &            &            &    11.841  &    11.267  &    10.965   &    9.960 &   3.99  \\
YSO-18    &      G018.2548-00.2441   &              &            &            &    12.808  &    11.200  &    10.534   &   10.170 &   3.31  \\
YSO-19    &      G018.2595-00.2488   &              &            &            &    13.088  &    12.392  &     9.999   &    8.284 &   2.80  \\
YSO-20    &      G018.2710-00.2657   &     16.190   &   14.493   &   13.778   &    11.499  &    10.849  &    10.284   &    9.601 &   3.96  \\
YSO-21    &      G018.2534-00.2492   &              &            &            &    13.645  &    11.846  &     8.622   &    6.972 &   2.18  \\
YSO-22    &      G018.2758-00.2636   &              &            &   12.501   &     9.470  &     8.050  &     6.964   &    6.279 &   2.91  \\
YSO-23    &      G018.2790-00.2652   &              &            &            &    13.475  &    11.528  &     9.935   &    8.880 &   5.16  \\
YSO-24    &      G018.2656-00.2405   &              &            &            &    10.827  &     8.850  &     7.235   &    6.160 &   2.97  \\
YSO-25    &      G018.2939-00.2528   &              &            &   14.025   &     9.671  &     8.224  &     7.351   &    6.945 &   4.16  \\
YSO-26    &      G018.2907-00.2629   &              &            &   13.910   &    11.870  &    11.183  &    10.614   &    9.913 &         \\
YSO-27    &      G018.2573-00.2506   &              &            &            &    13.125  &    11.227  &     9.054   &    8.372 &   2.73  \\
YSO-28    &      G018.2915-00.2689   &              &            &            &    13.322  &    11.806  &    10.490   &    9.663 &         \\
YSO-29    &      G018.2311-00.3150   &              &            &            &    11.914  &    11.406  &    10.533   &    9.611 &   3.41  \\
YSO-30    &      G018.2383-00.2910   &     14.385   &   13.076   &   11.883   &    10.262  &     9.702  &     8.984   &    7.592 &         \\
YSO-31    &      G018.2362-00.3032   &              &            &            &    13.021  &    11.723  &    10.807   &    9.904 &   2.76  \\
YSO-32    &      G018.2755-00.3118   &              &            &            &    13.188  &    12.867  &    10.458   &    8.923 &   1.24  \\
\hline
\end{tabular}
\tablefoot{The numbers of the YSOs correspond to the numeration in Fig. \ref{8} and Fig. \ref{9}.}
\end{table*}

Owing to the proper motion of ionizing star(s), they are not always
located in the geometrical center of the \HII\ region. But because
of the young ages of the IR dust bubble, the ionizing star(s) may be located
near the geometrical center and still lie inside the cavity created
by stellar wind. We can see that five O-type stars, \#1, \#4, \#7,
\#10, and \#13, are located in projection inside the cavity of the
20 cm radio continuum emission (see Fig. \ref{7}), they are probably
the exciting stars of N22.

\subsection{SED fitting and star formation}
\label{sec 4.4}

\begin{figure}[htb]
\tiny
\resizebox{\hsize}{!}{\includegraphics{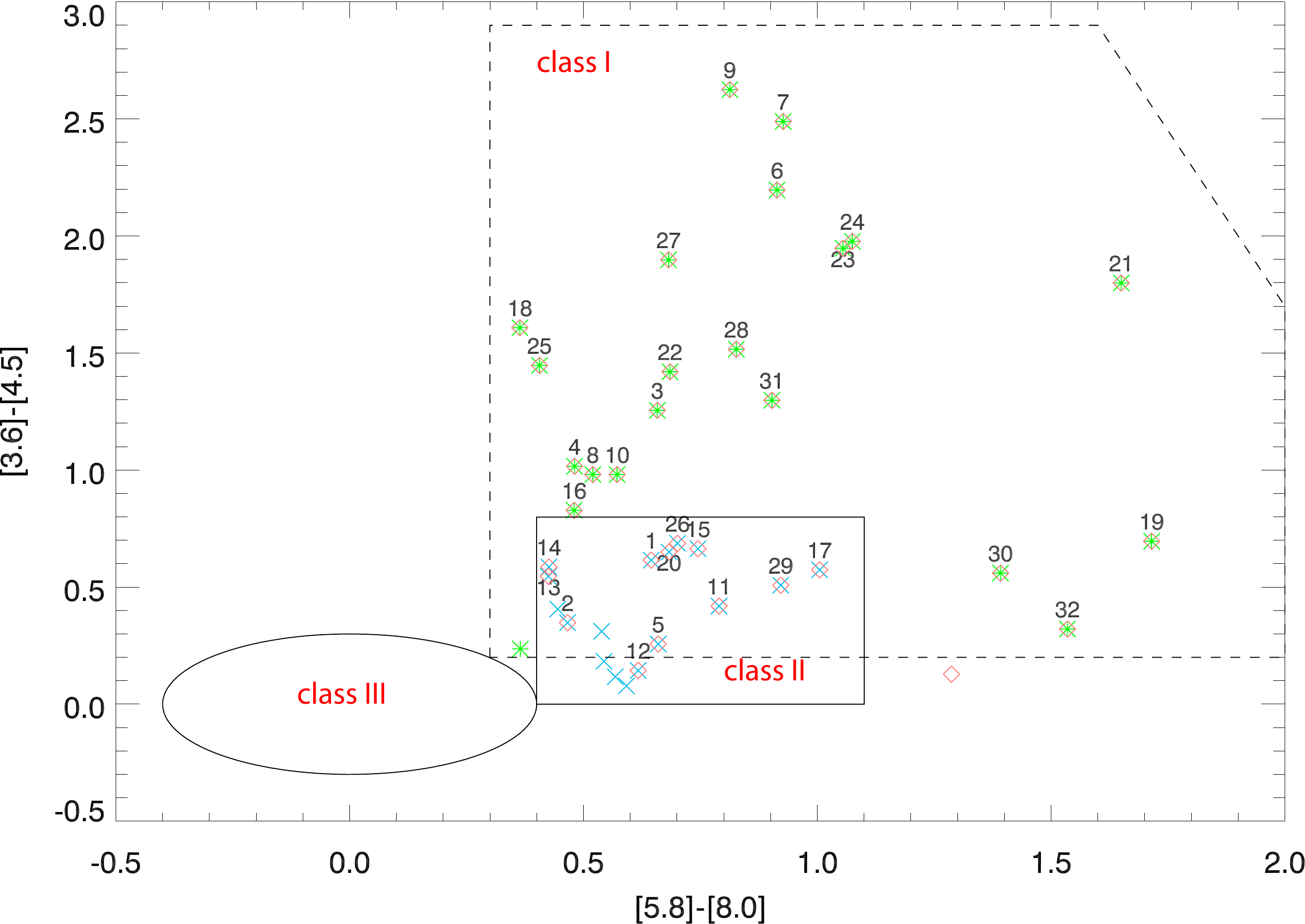}}
\caption{GLIMPSE color-color diagram [5.8]-[8.0] versus [3.6]-[4.5] for sources within a circle of 4.\m5 in radius centered at N22. We only considered sources with detection in the four {\it Spitzer}-IRAC bands. The regions indicate the stellar evolutionary stage (green star = class I, blue crossing = class II) as defined by Allen et al. (\cite{all04}) with the color criterion of $m_{4.5} - m_{8.0} > 1$ (red diamonds).}
\label{8}
\end{figure}

\begin{figure*}[htb]
\tiny
\centering
\includegraphics[width=17cm]{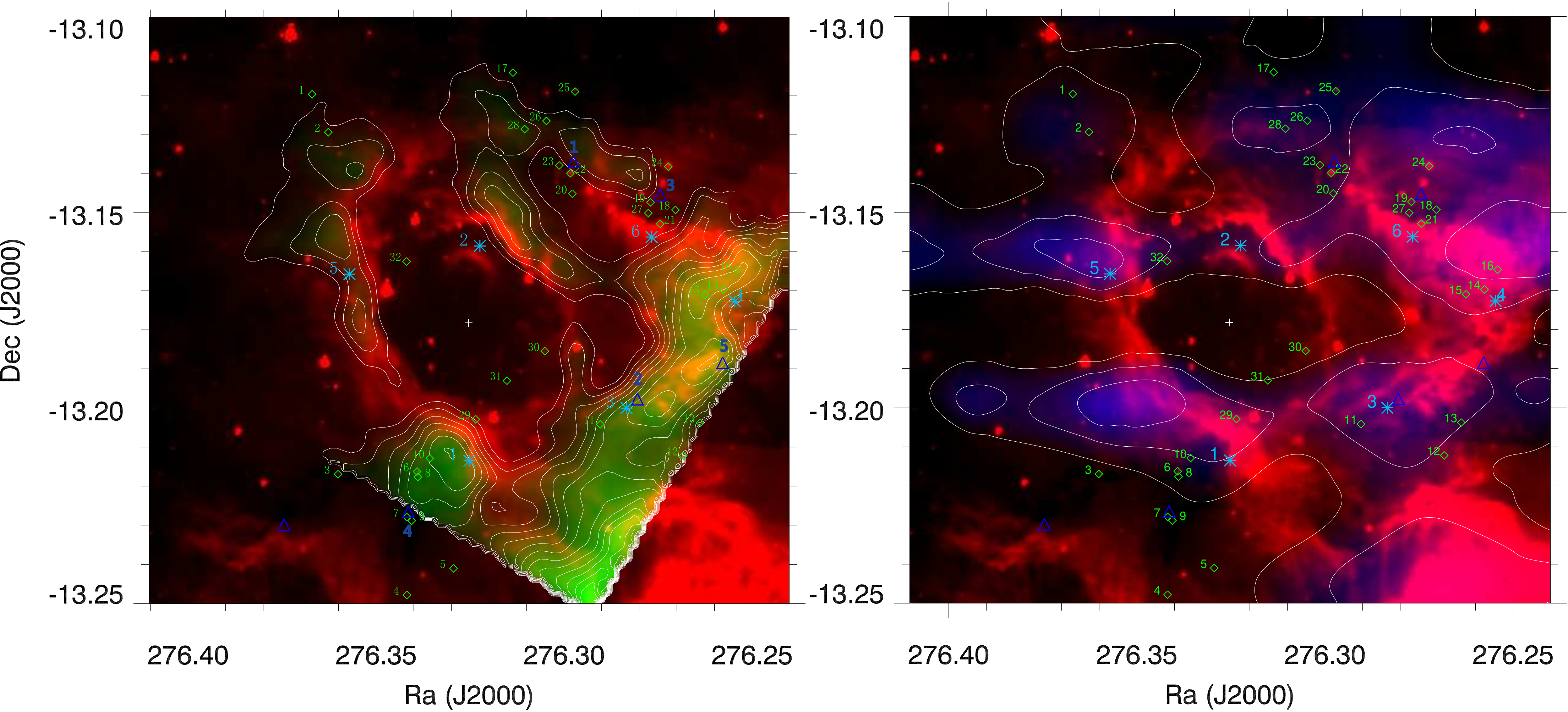}
\caption{Left: The background and contours are similar to Fig. \ref{4}.
Right: The background and contours are similar to Fig. \ref{6}.
The 32 sources of {\it Spitzer}-IRAC are symbolized with green diamonds.
The blue stars are FIS-AKARI point sources,
and triangles are BOLOCAM sources.
The white plus is the center of N22. }
\label{9}
\end{figure*}

Our analysis suggests that triggered star formation may take place
around N22. To search for YSO candidates within a circle of 4.\m5 in
radius centered at N22, we constructed a color-color diagram
[5.8]-[8.0] versus [3.6]-[4.5] with the sources that have fluxes in
the four {\it Spitzer}-IRAC bands. We used the photometric criteria
of Allen et al. (\cite{all04}) and the color criterion $m_{4.5} -
m_{8.0} > 1$ (Robitaille et al. \cite{rob08}) to identify class I and II
YSOs (see  Fig. \ref{8}). Finally, we selected 32 sources around N22
(see Fig. \ref{9}) and 10 of them, YSO-3, 4, 6, 9, 15, 19, 22, 24,
25, and 30, are cataloged by Robitaille et al. (\cite{rob08}) as
Galactic midplane {\it Spitzer} red sources. Robitaille et al.
(\cite{rob08}) pointed out that at most 0.4 \% of the intrinsically
red sources selected by the color criterion $m_{4.5} - m_{8.0} > 1$
are galaxies and AGNs. Hence, there is a low probability of finding
an extragalactic source in our small sample.

\begin{figure*}[htb]
\tiny
\centering
\includegraphics[width=17cm]{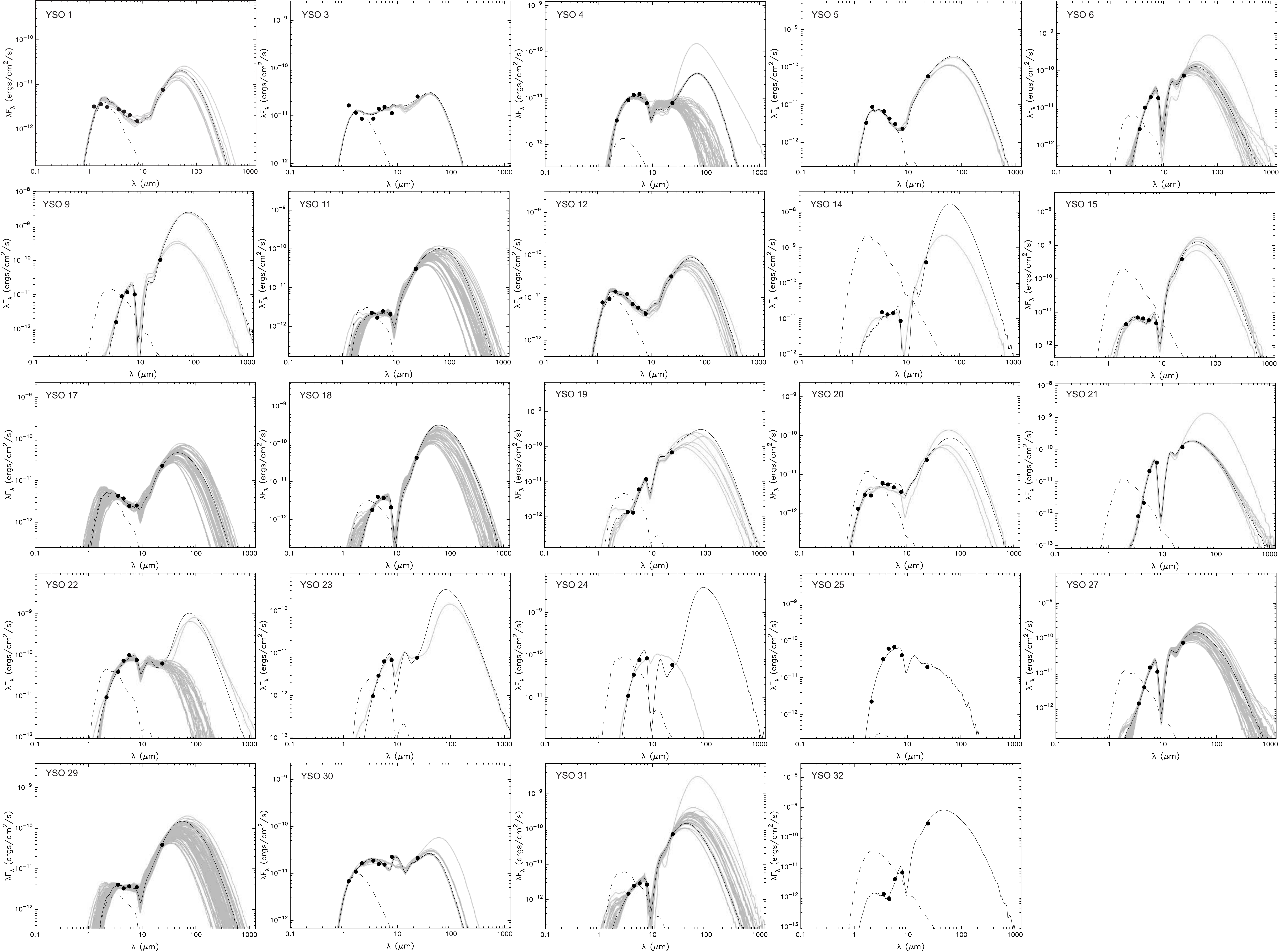}
\caption{SED of sources from which we obtained fluxes at 24 $\mu$m from the MIPS image.
The sources are numbered according to Figs. \ref{8} and \ref{9}.
In each panel, the black line shows the best fit, and the gray lines show subsequent good fits.
The dashed line shows the stellar photosphere corresponding to the central source of the best-fitting model,
as it would look without circumstellar dust. The points are the input fluxes.}
\label{10}
\end{figure*}

The contributions from the disk and envelope of YSOs in the spectral
energy distribution (SED) usually comes from long wavelengths, which
are longer than $\sim$ 10 microns and up to the millimeter region of
the electromagnetic spectrum. Therefor longer wavelength data are needed
to obtain a reliable SED for a YSO. We used the 'phot' tool of Aladin
(Bonnarel et al.~\cite{bon00}) to obtain the 24 $\mu$m fluxes of the
sources from {\it Spitzer}-MIPS image.
Using the conversion factor of 24 $\mu$m flux (7.17 Jy)
toward its zero magnitude (Engelbracht et al.~\cite{eng07}),
we calculated the 24 $\mu$m magnitudes.
The SED-fitting tool developed by
Robitaille et al. (\cite{rob07}) is available
online\footnote{http://caravan.astro.wisc.edu/protostars/}.
We used this tool to fit the SED of all YSO candidates and selected
the SED best-fit models according to the condition
$\chi^{2} - \chi_{best}^{2}~per~datapoint < 3$, where ~\cchib\ is the \cchi\ of
the YSO best-fit model. Finally, we obtained the magnitudes of 24
sources and fitted their SED; the fitting results are listed
in Table \ref{t3}. Fig. \ref{10} shows the SED of these sources.
Here we used the {\it (J-H)} vs. {\it (H-K)} color-color diagram
from 2MASS to derive the visual extinction of those sources;
the values are between 3 and 35 mag. Most of the values exceed 10 mag.
According to Neckel \& Klare (\cite{nec80}),
the visual extinction toward star-forming regions generally exceeds 10 magnitudes.
Finally, we obtained the extinction between 10 and 35 mag.

The SED fitting allows us to obtain the physical parameters of the
sources: the central source mass $M_\star$, the disk mass $M_{\rm
disk}$, the envelope mass $M_{\rm env}$, and the envelope accretion
rate $\dot{M}_{\rm env}$. According to these parameters, Robitaille
et al. (\cite{rob06}) classified YSOs into three stages: stage 0/I
objects have significant infalling envelopes and possibly disks,
they have $\dot{M}_{\rm env}/M_\star>10^{-6}$ yr$^{-1}$; stage II
objects have optically thick disks (and possible remains of a
tenuous infalling envelope), they have $M_{\rm disk}/M_\star>10^{-6}$ and
$\dot{M}_{\rm env}/M_\star<10^{-6}$ yr$^{-1}$; and stage III objects
have optically thin disks, they have $M_{\rm disk}/M_\star<10^{-6}$
and $\dot{M}_{\rm env}/M_\star<10^{-6}$ yr$^{-1}$.
From the SEDs of the sources (see Fig. \ref{10}),
we found that YSO-3, 25 and 30 are stage II objects, YSO-4, 22 and 24 could be
stage I and II objects, and we suggest that the remaining 18 sources are mainly stage I sources.
We have to note that due to the absence of longer wavelengths data,
the results of the stage I sources may have some uncertainty.

\begin{table}
\tiny
\caption{The fluxes for the six FIS-AKARI point sources}
\label{t5}
\centering
\begin{tabular}{cccccc}
\hline\hline
Source &  Name  &  $S65$ & $S90$ & $S140$ & $S160$ \\
       &        &  (Jy)  & (Jy)  &  (Jy)  &  (Jy)  \\
\hline
1	&	1825180-131240	& 142.3	&	 132.6 &  360.0	& 291.0	\\
2	&	1825174-130920	& 231.5	&	  90.7 &        &       \\
3	&	1825080-131151	& 125.4	&	  73.8 &  	    & 125.6	\\
4	&	1825011-131012	& 290.0	&	 182.8 &  531.7	& 431.5 \\
5	&	1825257-130946	&  79.9	&	  97.5 &  246.6	& 687.0	\\
6	&	1825064-130912	& 226.5	&	 170.2 &  488.9	& 721.5	\\	

\hline
\end{tabular}
\tablefoot{The numbers of the YSOs (FIS-AKARI point sources) correspond to the numeration in Fig. \ref{9}.}
\end{table}

\begin{figure}[htb]
\tiny
\resizebox{\hsize}{!}{\includegraphics{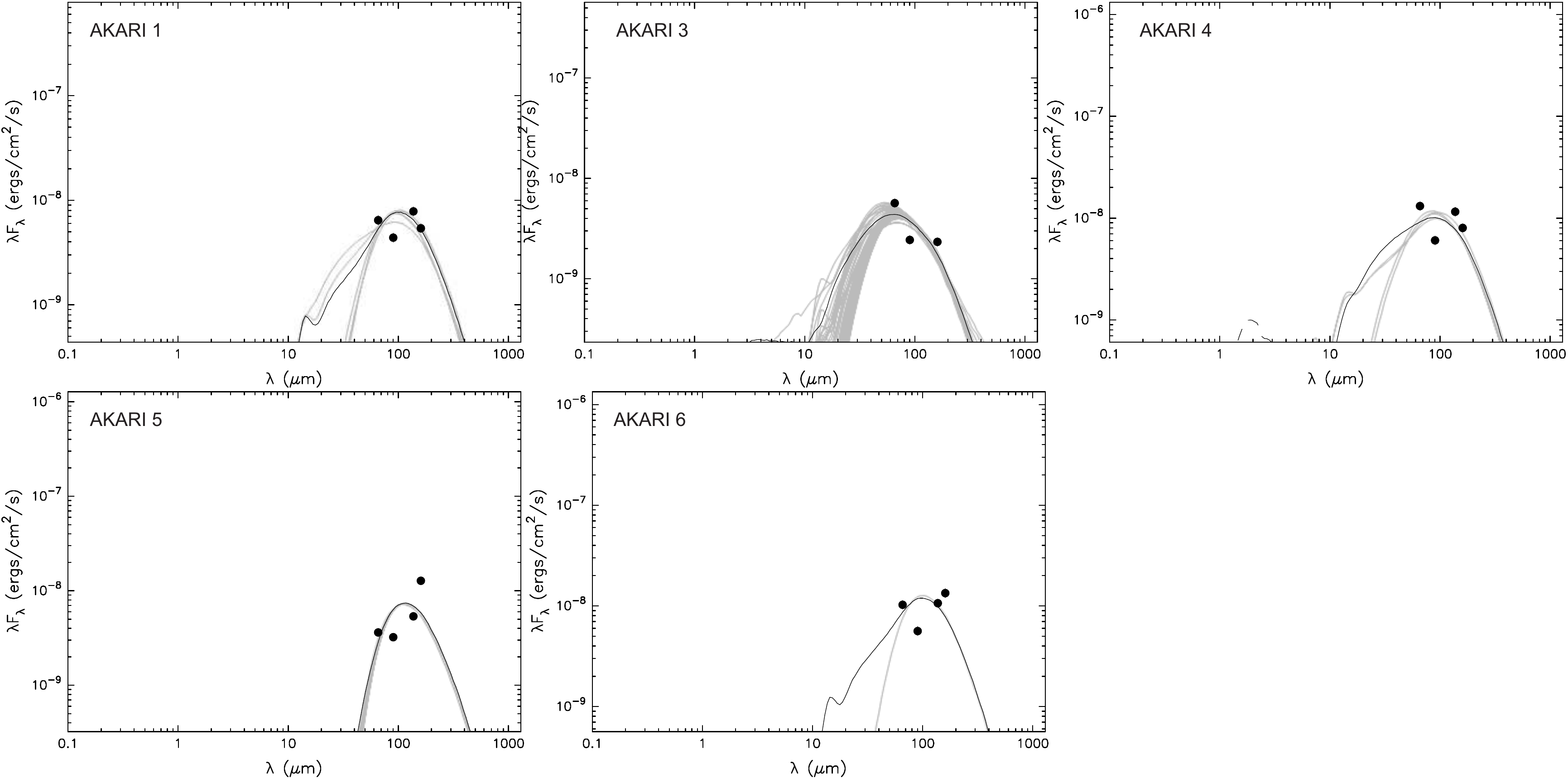}}
\caption{SED of sources from the five FIS-AKARI sources around N22.
The sources are numbered according to Table \ref{t6}.}
\label{11}
\end{figure}

We also found six FIS-AKARI point sources around N22,
whose fluxes are listed in Cols. 3-6 of Table \ref{t5}.
Here, we used the Robitaille on-line tool to fit the FIS-AKARI point source fluxes
by specifying 'monochromatic' instead of 'broad/narrow-band'.
We did not fit source 2, because only two bands are detected.
Finally, the SED fitting of the five FIS-AKARI sources indicates that they are all
stage 0/I objects with masses ranging from 6.1 to 14.7 \msol, their ages are
about $10^3 \sim 10^5$ yr (see Table \ref{t6}). Fig. \ref{11} shows the SEDs of the five FIS-AKARI sources.

\begin{table*}
\tiny
\caption{Parameters derived from the SED fitting of five YSOs (FIS-AKARI point sources)}
\label{t6}
\centering
\begin{tabular}{ccccccccc}
\hline\hline
source & \cchib/$N$  & $n$  & $M_\star$  & $Age$  & $M_{disk}$  & $M_{env}$& $\dot{M}_{env}$  & stage  \\
     &             &      &  (\msol)   & (yr)   &  (\msol)    & (\msol)  &(\msol/yr)        &          \\
\hline
 1     &       18.66   &     1  &   14.7          &   $                     3.2 \times  10^{3}      $ &   $                     3.0 \times 10^{-2}   $   &  $                      4.4 \times  10^{3}   $   &   $                      7.2 \times 10^{-3}  $  &     0/I  \\
 3     &      338.00   &     2  &    6.1- 6.6     &   $  1.9 \times  10^{4}-1.0 \times  10^{5}      $ &   $  8.7 \times 10^{-4}-1.9 \times 10^{-3}   $   &  $   5.1 \times  10^{1}-7.1 \times  10^{1}   $   &   $   5.6 \times 10^{-4}-2.2 \times 10^{-3}  $  &     0/I  \\
 4     &       32.38   &     6  &   11.4-11.4     &   $  4.4 \times  10^{3}-4.4 \times  10^{3}      $ &   $  2.3 \times 10^{-1}-2.3 \times 10^{-1}   $   &  $   3.7 \times  10^{3}-3.7 \times  10^{3}   $   &   $   5.4 \times 10^{-3}-5.4 \times 10^{-3}  $  &     0/I  \\
 5     &       10.96   &    63  &    6.4-11.3     &   $  1.3 \times  10^{3}-1.1 \times  10^{5}      $ &   $  0.0               -1.3                  $   &  $   2.0 \times  10^{2}-1.3 \times  10^{3}   $   &   $   5.3 \times 10^{-4}-2.1 \times 10^{-3}  $  &     0/I  \\
 6     &        8.99   &    29  &   10.7-14.0     &   $  2.0 \times  10^{3}-3.9 \times  10^{4}      $ &   $  0.0               -3.1 \times 10^{-1}   $   &  $   1.3 \times  10^{3}-3.7 \times  10^{3}   $   &   $   2.5 \times 10^{-3}-5.4 \times 10^{-3}  $  &     0/I  \\
\hline
\end{tabular}
\tablefoot{The numbers of the YSOs (FIS-AKARI point sources) correspond to the numeration in Fig. \ref{9}.}
\end{table*}

Most of the 24 sources and the five FIS-AKARI point sources lie close to
the dense cores and the active region around the N22 (see Fig. \ref{9}).
Thus star formation is indeed active around N22. The formation of these
YSOs may have been triggered by the expanding \HII\ region.

From Fig. \ref{9} ({\it left}), the five BOLOCAM sources can been seen across the borders of N22.
Considering the gas-to-dust ratio of 100 (Enoch et al.~\cite{eno06}),
we estimated the BOLOCAM source masses following the method of Rosolowsky et al. (\cite{ros10}):
$$M=\frac{13.1}{100} (\frac{D}{1~\rm kpc})^2 (\frac{S_{\nu}}{1~\rm Jy}) [\frac{\rm exp(13.0~\rm K/{\it T})-1}{\rm exp(13.0/20)-1}]~[M_\odot],$$
where $S_{\nu}$ is the total flux density of the BOLOCAM source in the catalog,
dust temperature $T=20$ K and distance $D=4.1$ kpc.
We obtained dust masses of the five BOLOCAM source (\#1 to \#5) of
7.1 \msol, 0.8 \msol, 18.7 \msol, 7.7 \msol, and 0.4 \msol, respectively.

\section{Collect-and-collapse scenario ?}
\label{sec 5}

We estimated the age of the \HII\ region and the fragmentation time
to examine whether the "collect-and-collapse" mechanism is
responsible for the star formation around the bubble N22.

To obtain the age of the \HII\ region, we used the model described by
Dyson \& Williams (\cite{dys80}) with a given radius $R$ as
$$t(R)=\frac{4~R_s}{7~c_s}\left[\left(\frac{R}{R_s}\right)^{7/4}-1\right],$$
where $c_s$ is the sound velocity in the ionized gas ($c_s$=10 km
s$^{-1}$) and $R_s$ is the radius of the Str\"omgren sphere given
by $R_s=(3N_{\rm Lyc}/4\pi n_0^2\alpha_B)^{1/3}$, where $N_{\rm
Lyc}$ is the number of ionizing photons emitted by the star per
second, $n_0$ is the original ambient density, and $\alpha_B$=2.6
$\times 10^{-13}$ cm$^3$ s$^{-1}$ is the hydrogen recombination
coefficient to all levels above the ground level. Here we adopted a
Lyman continuum photon flux of about $4.2 \sim 9.4 \times 10^{49}$
ph s$^{-1}$ (the total Lyman continuum photon flux of exciting
stars, \#1, \#4, \#7, \#10, and \#13, see Sec. 4.3), a radius of
$\sim$ 1.77 pc, and an original ambient density of $\sim 1.0 \times
10^3$ cm$^{-3}$. Finally, we derived a dynamical age of between 0.06
and 0.15 Myr for N22. To roughly estimate the original ambient
density, we distributed the total mass of most dense cores, $\sim 1.2
\times 10^{4}$ \msol\ (exclude core 1 and core 4), over a sphere
with a radius of 2 pc.

As mentioned above, several pieces of evidence indicate
that the expansion of N22 collects the gas at its periphery.
To determine whether the fragmentation could occur around N22,
we estimated the fragmentation time of the collected layer
according to the theoretical models of Whitworth et al. (\cite{whi94b}):
$$t_{\rm fragment}=1.56 (\frac{\alpha_s}{0.2})^{7/11}(\frac{N_{\rm Lyc}}{10^{49}})^{-1/11} (\frac{n_0}{10^3})^{-5/11}~[\rm Myr].$$
Using a turbulent velocity $\alpha_s$ ranging between 0.2 and 0.6 \ks\ (Whitworth et al.~\cite{whi94b}),
we find that the fragmentation of the collected layer should occur between 1.36 and 2.93 Myr after its
formation, which is later than the dynamical age.
The fragmentation time is inferred by considering the uncertainty in the
total Lyman continuum photon flux and turbulent velocity.
Hence, the "collect-and-collapse" mechanism seems not to be responsible for
the star formation activities around N22.
Other processes such as the radiation-driven compression of pre-existing dense clumps
(Deharveng et al.~\cite{deh10}) may operate here.
From Figs. \ref{1} ({\it right}) and \ref{9} ({\it left}),
we state that the compression of pre-existing dense clumps
should be taking place mainly at the borders of IRDC-A and IRDC-B,
which are interacting with N22.
The YSOs (YSO-6, 7, 8, 9, 10, and AKARI-1) related to the IRDC-B
are very likely formed by this process.

\section{Summary}
\label{sec 6}

Using multiwavelength surveys and archival data, we have studied the ISM around the bubble N22.
The main results can be summarized as follows:

(1) The PAH emission around N22 is detected at 8 $\mu$m and the radius is
about 1.77 pc. The 24 $\mu$m emission reveals hot
dust in the interior of the \HII\ region. The 20 cm emission is
bounded by 8 $\mu$m emission and shows a shell morphology. The 20 cm
emission is distributed mainly toward the north of the shell with two
peaks. A cavity can be clearly seen at 20 cm emission, which may be
created by the exciting-star(s) of N22.

(2) A molecular shell composed of several clumps is distributed around
the \HII\ region. Among the clumps, core 1 and core 4 are associated with IRDC-A and IRDC-B,
which appear as dark extinction features against the Galactic background at mid-infrared wavelengths.
The molecular shell has a total mass of about 21 000 \msol.

(3) The integrated KOSMA \2 line intensity ratios $R_{I_{{\rm
CO(3-2)}}/I_{{\rm CO(2-1)}}}$ around N22 are measured at between 0.7 and 1.14.
The higher line ratios (higher than 0.8) imply that
shocks have driven into the MCs, this suggest that the expanding
\HII\ region is interacting with the surrounding MCs.

(4) We found 11 O-type star inside the \HII\ region, five
of which are located in projection inside the cavity of the 20 cm
radio continuum emission. We suggest that the five O-type stars
may be a cluster exciting the bubble N22.

(5) We discovered 24 YSOs that are very likely embedded in the molecular shell.
Of these three YSOs are stage I-II objects, three YSOs are stage-II objects,
and 18 YSOs are suggested to stage-I objects.
Most of these 24 YSOs lie close to the dense cores and the active regions around the N22.
We suggest that the formation of these YSOs are probably be triggered by the expanding \HII\ region.

(6) We also fitted the SED of five FIS-AKARI point sources that are located close to the dense cores.
They are all massive stars in stage 0/I.

(7) By comparing the dynamical age of N22 and the
fragmentation time of the molecular shell surrounding N22, we
suggest that the triggered star formation mechanism "radiation-driven
compression of pre-existing dense clumps" may work here.
We state that IRDC-A and IRDC-B, which are the pre-existing dense clumps, are interacting with N22,
and conclude that the YSOs related to the IRDC-B are very likely formed by this process.

\section*{Acknowledgments}

This work was funded by The National Natural Science Foundation of China under grant 10778703 and was partly supported by China Ministry of Science and Technology under State Key Development Program for Basic Research (2012CB821800) and The National Natural Science Foundation of China under grant 10873025.

\bibliographystyle{aa}  
\bibliography{biblio}
\IfFileExists{\jobname.bbl}{}
{\typeout{}
\typeout{****************************************************}
\typeout{****************************************************}
\typeout{** Please run "bibtex \jobname" to optain}
\typeout{** the bibliography and then re-run LaTeX}
\typeout{** twice to fix the references!}
\typeout{****************************************************}
\typeout{****************************************************}
\typeout{}
}



\label{lastpage}

\end{document}